\documentclass[a4paper,UKenglish,cleveref]{lipics-v2019}
\title{List $k$-Colouring $P_t$-Free Graphs: a Mim-width Perspective}

\author{Nick Brettell}{School of Mathematics and Statistics, Victoria University of Wellington, New Zealand}{nick.brettell@vuw.ac.nz}{https://orcid.org/0000-0002-1136-418X}{}

\author{Jake Horsfield}{School of Computing, University of Leeds, Leeds,
UK}{sc15jh@leeds.ac.uk}{https://orcid.org/0000-0002-4388-5123}{}

\author{Andrea Munaro}{School of Mathematics and Physics, Queen's University Belfast, UK}{a.munaro@qub.ac.uk}{https://orcid.org/0000-0003-1509-8832}{}

\author{Dani\"el Paulusma}{Department of Computer Science, Durham University, UK}{daniel.paulusma@durham.ac.uk}{https://orcid.org/0000-0001-5945-9287}{}

\authorrunning{N. Brettell, J. Horsfield, A. Munaro, and D. Paulusma}

\Copyright{Nick Brettell, Jake Horsfield, Andrea Munaro, and Dani\"el Paulusma}



\funding{The research in this paper received support from the Leverhulme Trust (RPG-2016-258).  The first author was supported by a Rutherford Foundation Postdoctoral Fellowship from Royal Society Te Apārangi.}

\acknowledgements{}

\nolinenumbers

\hideLIPIcs

\usepackage{amsmath,amssymb,amsthm,xspace,xcolor,tikz,graphicx,graphics,color,tkz-graph}
\usetikzlibrary{fit,automata,arrows}

\newcommand{\NP}{{\sf NP}}
\newcommand{\XP}{{\sf XP}}

\newcommand{\cutmim}{\mathrm{cutmim}}
\newcommand{\mimw}{\mathrm{mimw}}

\begin{document}

\maketitle

\begin{abstract} 
A colouring of a graph $G=(V,E)$ is a mapping $c\colon V\to \{1,2,\ldots\}$ such that $c(u)\neq c(v)$ for every two adjacent vertices $u$ and $v$ of $G$.
The {\sc List $k$-Colouring} problem is to decide whether a graph $G=(V,E)$ with a list $L(u)\subseteq \{1,\ldots,k\}$ for each $u\in V$ has a colouring $c$ such that $c(u)\in L(u)$ for every $u\in V$.  
Let $P_t$ be the path on $t$ vertices and let $K_{1,s}^1$ be the graph obtained from the $(s+1)$-vertex star $K_{1,s}$ by subdividing each of its edges exactly once.

\medskip
\noindent
Recently, Chudnovsky, Spirkl and Zhong (DM 2020) proved that {\sc List $3$-Colouring} is polynomial-time solvable for $(K_{1,s}^1,P_t)$-free graphs for every $t\geq 1$ and $s\geq 1$.
We generalize their result to {\sc List $k$-Colouring} for every $k\geq 1$.
Our result also generalizes the known result that for every $k\geq 1$ and $s\geq 0$, {\sc List $k$-Colouring} is polynomial-time solvable for $(sP_1+P_5)$-free graphs, which was proven for $s=0$ by Ho\`ang, Kami\'nski, Lozin, Sawada, and Shu (Algorithmica 2010) and for every $s\geq 1$ by Couturier, Golovach, Kratsch and Paulusma (Algorithmica 2015).

\medskip
\noindent
We show our result by proving boundedness of an underlying width parameter. Namely, we show 
that for every $k\geq 1$, $s\geq 1$, $t\geq 1$, the class of $(K_k,K_{1,s}^1,P_t)$-free graphs has bounded mim-width and that a corresponding branch decomposition is ``quickly computable'' for these graphs.
\end{abstract}

\section{Introduction}

\begin{figure}[h]
  \centering
\begin{minipage}[c]{0.22\textwidth}
\begin{tikzpicture}[scale=0.88] \draw(-1,2)--(-1,0);
\draw(-1,2)--(-1,0);
\draw(-1,0)--(-1,-2);
\draw(-1,-2)--(1,-2);
\draw(1,-2)--(1,0);
\draw(1,0)--(1,2);
\draw(-1,0)--(1,-2);
\draw(-1,-2)--(1,0);
\draw[fill=white] (1,2) circle [radius=2pt] (1,0) circle [radius=2pt] (1,-2) circle [radius=2pt]
(-1,2) circle [radius=2pt] (-1,0) circle [radius=2pt] (-1,-2) circle [radius=2pt];
\node[right] at (1,-2) {$v_6$}; \node[right] at (1,0) {$v_5$}; \node[right] at (1,2) {$v_4$};
\node[left] at (-1,2) {$v_3$}; \node[left] at (-1,0) {$v_2$}; \node[left] at (-1,-2) {$v_1$};
\end{tikzpicture}
\end{minipage}
\qquad
\begin{minipage}[c]{0.3\textwidth}
\begin{tikzpicture}[scale=0.88] \draw[color=black, fill=gray!5] (1.1,-2) ellipse (0.6cm and 1cm);
\draw[color=black, fill=gray!5] (1.1,1) ellipse (0.9cm and 1.9cm); \draw[very thick] (-1,0)--(0,-1.8);
\draw(1,2.5)--(0,1.8)--(-1,0)--(0,0)--(1,0.5)(1,-2.5)--(0,-1.8)--(1,-1.5) (1,-0.5)--(0,0) (1,1.5)--(0,1.8);
\draw[fill=white] (1,2.5) circle [radius=2pt] (1,1.5) circle [radius=2pt] (1,0.5) circle [radius=2pt]
(1,-2.5) circle [radius=2pt] (1,-1.5) circle [radius=2pt] (1,-0.5) circle [radius=2pt]
(0,1.8) circle [radius=2pt](0,-1.8) circle [radius=2pt](0,0) circle [radius=2pt] (-1,0) circle [radius=2pt];
\node[right] at (1,2.5) {$v_4$}; \node[right] at (1,1.5) {$v_5$}; \node[right] at (1,0.5) {$v_6$};
\node[right] at (1,-0.5) {$v_1$}; \node[right] at (1,-1.5) {$v_2$}; \node[right] at (1,-2.5) {$v_3$};
\node[right] at (-0.5,-0.8) {$e$}; \node[right] at (1.7,-2) {$L_e$};
\node[right] at (2,1) {$\overline{L_e}$}; \node[above] at (-1,1.5) {$(T,\delta)$};
\end{tikzpicture}
\end{minipage}
\qquad
\begin{minipage}[c]{0.3\textwidth}
\begin{tikzpicture}[scale=0.88] \draw[color=black, fill=gray!5] (-1,0) ellipse (0.7cm and 1.7cm);
\draw[color=black, fill=gray!5] (1,0) ellipse (1.1cm and 2.8cm);
\draw[very thick] (1,-0.66)--(-1,-1.2); \draw (1,-2)--(-1,-1.2);
\draw[dotted] (-1,1.2)--(-1,-1.2) (1,0.66)--(1,-0.66) (1,2)--(1,0.66) (1,-0.66)--(1,-2) (1,0.66)to[out=300,in=60](1,-2);
\draw[fill=white] (1,2) circle [radius=2pt] (1,0.66) circle [radius=2pt] (1,-0.66) circle [radius=2pt]
(1,-2) circle [radius=2pt] (-1,1.2) circle [radius=2pt] (-1,-1.2) circle [radius=2pt];
\node[below left] at (-1,1.2) {$v_3$}; \node[above left] at (-1,-1.2) {$v_2$}; \node[above left] at (1,2) {$v_4$};
\node[left] at (1,0.66) {$v_5$}; \node[above left] at (1,-0.66) {$v_6$}; \node[below left] at (1,-2) {$v_1$};
\node[below] at (-1,-2) {$A_e$}; \node[below] at (1,-2.9) {$\overline{A_e}$};
\end{tikzpicture}
\end{minipage}
\caption{An example of a graph $G$ with a branch decomposition $(T,\delta)$. The partition $(A_{e}, \overline{A_{e}})$ of $V(G)$ in the rightmost figure witnesses that $\mimw_{G}(T, \delta)\geq 1$. It can be easily seen that $\mimw_{G}(T, \delta)\leq 1$ and so $\mimw(G)=1$.}\label{l-exx}
\end{figure}
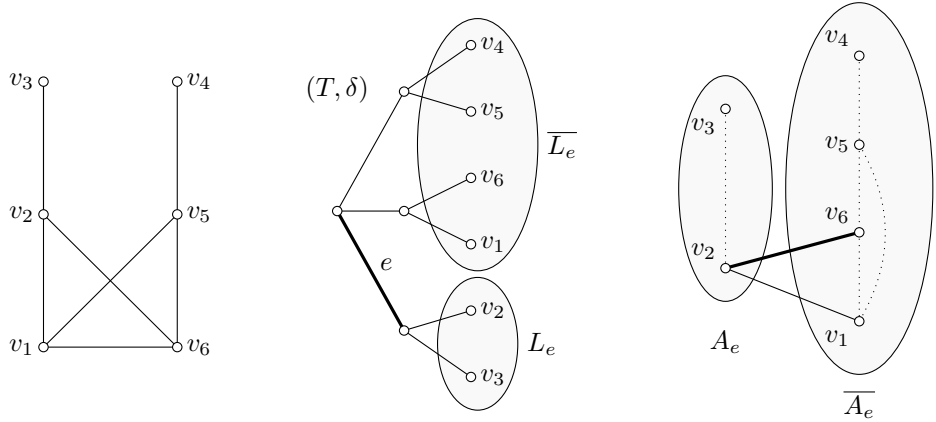

Width parameters play an important role in algorithmic graph theory, as evidenced by various surveys~\cite{DJP19,Gu17,HOSG08,KLM09,Ou17}.
A graph class ${\cal G}$ has {\it bounded} width, for some width parameter, if there exists a constant~$c$ such that every graph in ${\cal G}$ has width at most~$c$.
Mim-width is a relatively young width parameter that was introduced by Vatshelle~\cite{Va12}.
It is defined as follows. A \textit{branch decomposition} for a graph~$G$ is a pair $(T, \delta)$, where $T$ is a subcubic tree and $\delta$ is a bijection from~$V(G)$ to the leaves of $T$.  
Every edge $e \in E(T)$ partitions the leaves of $T$ into two classes, $L_e$ and $\overline{L_e}$, depending on which component of $T-e$ they belong to.
Hence, $e$ induces a partition $(A_e, \overline{A_e})$ of $V(G)$, where $\delta(A_e) = L_e$ and $\delta(\overline{A_e}) = \overline{L_e}$.
We let $G[A_e,\overline{A_e}]$ be the bipartite subgraph of $G$ induced by the edges with one end-vertex in $A_e$ and the other in $\overline{A_e}$.
A matching $F \subseteq E(G)$ of $G$ is {\it induced} if there is no edge in $G$ between vertices of different edges of $F$.
We let $\cutmim_{G}(A_{e}, \overline{A_{e}})$ be the size of a maximum induced matching in $G[A_{e}, \overline{A_{e}}]$.
The \emph{mim-width} $\mimw_{G}(T, \delta)$ of $(T, \delta)$ is the maximum value of $\cutmim_{G}(A_{e}, \overline{A_{e}})$ over all edges $e\in E(T)$. The \emph{mim-width} $\mimw(G)$ of $G$ is the minimum value of $\mimw_{G}(T, \delta)$ over all branch decompositions $(T, \delta)$ for $G$. See Figure~\ref{l-exx} for an example.

Vatshelle~\cite{Va12} proved that every class of bounded clique-width, or equivalently, bounded boolean-width, module-width, NLC-width or rank-width, has bounded mim-width, and that the converse is not true. That is, he proved that there exist graph classes of bounded mim-width that have unbounded clique-width. 
This means that proving that a problem is polynomial-time solvable for graph classes of bounded mim-width
yields more tractable graph
classes than doing this for clique-width. Hence, mim-width has greater {\it modeling power} than clique-width.

However, the {\it trade-off} is that fewer problems admit such an algorithm, as we explain below by means of a relevant example, namely the classical {\sc Colouring} problem.
Moreover, computing mim-width is \NP-hard~\cite{SV16} and it is not possible to approximate in polynomial time the mim-width of a graph within a constant factor unless $\mathsf{NP} = \mathsf{ZPP}$~\cite{SV16}.
It remains a challenging open problem to develop a polynomial-time algorithm for computing a branch decomposition with mim-width $f(k)$ for a graph with mim-width~$k$.
However, the latter has been shown possible for special graph classes~${\cal G}$.
In such a case, we say that the mim-width of~${\cal G}$ is {\it quickly computable}.
We can then develop a polynomial-time algorithm for the problem of interest via dynamic programming over the computed branch decomposition.
We refer to~\cite{BV13,BK19,BPT19,BHMPP,BMP,BTV13,GMR20,JKST19,JKT,JKT19,KKST17} for a wide range of examples of graph classes and problems for which such dynamic programming algorithms have been obtained. 

\smallskip
As mentioned, in this paper we focus on Graph Colouring, a central problem in Discrete Mathematics, Theoretical Computer Science and beyond.
A {\em colouring} of a graph $G=(V,E)$ is a mapping $c\colon V\rightarrow\{1,2,\ldots \}$ that gives each vertex~$u\in V$ a {\it colour} $c(u)$ in such a way that, for every two adjacent vertices $u$ and $v$, we have that $c(u)\neq c(v)$. If for every $u\in V$ we have $c(u)\in \{1,\ldots,k\}$, then we say that $c$ is a {\it $k$-colouring} of $G$.
The {\sc Colouring} problem is to decide whether a given graph $G$ has a $k$-colouring for some given integer $k\geq 1$. If $k$ is {\it fixed}, that is, not part of the input, we call this the $k$-{\sc Colouring} problem.  A classical result of Lov\'asz~\cite{Lo73} states that $k$-{\sc Colouring} is \NP-complete even if $k=3$.

The {\sc Colouring} problem is an example of a problem that distinguishes between classes of bounded mim-width and bounded clique-width: it is polynomial-time solvable for every graph class of bounded clique-width~\cite{KR03} but \NP-complete for circular-arc graphs~\cite{GJMP80}, a class of graphs of mim-width at most~$2$ and for which mim-width is quickly computable~\cite{BV13}. When we fix $k$, we no longer have this distinction, as
$k$-{\sc Colouring}, for every fixed integer $k\geq 1$, is polynomial-time solvable for a graph class whose mim-width is bounded and quickly computable~\cite{BTV13}.

We consider the following generalization of {\sc $k$-Colouring}.
For an integer $k\geq 1$, a  {\it $k$-list assignment} of a graph
$G=(V,E)$ is a function $L$ that assigns each vertex $u\in V$ a {\it list} $L(u)\subseteq \{1,2,\ldots,k\}$ of {\it admissible} colours for $u$. A colouring $c$ of $G$ {\it respects} $L$ if  $c(u)\in L(u)$ for every $u\in V$. For a fixed integer~$k\geq 1$, the {\sc List $k$-Colouring} problem is to decide whether a given graph~$G$ with a $k$-list assignment $L$ admits a colouring that respects $L$. 
Note that for $k_1\leq k_2$, {\sc List $k_1$-Colouring} is a special case of {\sc List $k_2$-Colouring} and that by setting
$L(u)=\{1,\ldots,k\}$ for every $u\in V$, we obtain the {\sc $k$-Colouring} problem.

Given an instance ($G,L)$ of {\sc List $k$-Colouring}, one can construct an equivalent instance $G'$ of {\sc $k$-Colouring} by adding a clique on new vertices $u_1,\ldots,u_k$ to $G$ and adding an edge between $u_i$ and $v\in V(G)$ if and only if $i\notin L(u)$ (see, for example,~\cite{LMS18}).
Kwon~\cite{Kw20} observed that $\mimw(G')\leq \mimw(G)+k$ and thus, as $k$-{\sc Colouring} is polynomial-time solvable for graph classes whose mim-width is bounded and quickly computable~\cite{BTV13}, for every fixed integer $k\geq 1$, this leads to the following:

\begin{theorem}[\cite{Kw20}]\label{t-lc}
For every $k\geq 1$, {\sc List $k$-Colouring} is polynomial-time solvable for a graph class whose mim-width is bounded and quickly computable.
\end{theorem}

\noindent
In this paper we show that a number of known polynomial-time results for {\sc List $k$-Colouring} on special graph classes can be obtained, and strengthened, by applying Theorem~\ref{t-lc}. 

The classes that we consider belong to the framework of hereditary graph classes. 
A graph class is {\it hereditary} if it is closed under vertex deletion. It is well known and not difficult to see that hereditary graph classes are exactly those classes characterized by a (unique) set ${\cal F}$ of minimal forbidden induced subgraphs. If $|{\cal F}|=1$ or $|{\cal F}|=2$, we say that the hereditary graph class is {\it monogenic} or {\it bigenic}, respectively.
In a recent study~\cite{BHMPP}, boundedness or unboundedness of mim-width has been determined for all monogenic classes and a large number of bigenic classes. These results imply that a monogenic graph class has bounded mim-width if and only if it has bounded clique-width~\cite{BHMPP} but that this equivalence does not always hold for bigenic graph classes. As we focus on hereditary graph classes, our work can be seen as a continuation of the research in~\cite{BHMPP}.

\subsection*{Related Work}

We first need to introduce some more terminology. A graph $G$ is {\it $H$-free}, for some graph~$H$, if it contains no {\it induced} subgraph isomorphic to~$H$, that is, we cannot modify $G$ into $H$ by a sequence of vertex deletions.
For a set of graphs $\{H_1,\ldots,H_p\}$, a graph is {\it $(H_1,\ldots,H_p)$-free} if it is $H_i$-free for every $i\in \{1,\ldots,p\}$. 
We denote the {\it disjoint union} of two graphs $G_1$ and $G_2$ by $G_1+G_2=(V(G_1)\cup V(G_2), E(G_1)\cup E(G_2))$. We let $P_r$ and $K_r$ denote the path and complete graph on $r$ vertices, respectively.

The complexity of {\sc Colouring} for $H$-free graphs has been settled for every graph $H$~\cite{KKTW01}, but there are still 
infinitely many open cases for {\sc $k$-Colouring} restricted to $H$-free graphs when $H$ is a {\it linear forest}, that is, 
a disjoint union of paths. We refer to~\cite{GJPS17} for a survey and to~\cite{CHSZ18,CSZ,KMMNPS18} for updated summaries. 
In particular, Ho\`ang et al.~\cite{HKLSS10} proved that for every integer $k\geq 1$, {\sc $k$-Colouring} is polynomial-time solvable for $P_5$-free graphs. Their proof is in fact a proof for {\sc List $k$-Colouring}.
The result of~\cite{HKLSS10} was generalized by Couturier et al.~\cite{CGKP15} as follows:

\begin{theorem}[\cite{CGKP15}]\label{t-known}
For every $k\geq 1$ and $s\geq 0$, {\sc List $k$-Colouring} is polynomial-time solvable for $(sP_1+P_5)$-free graphs. 
\end{theorem}

\noindent
For $r\geq 1$ and $s\geq 1$, we let $K_{r,s}$ denote the complete bipartite graph with partition classes of size $r$ and $s$. The graph $K_{1,s}$ is also known as the $(s+1)$-vertex star.
The {\it $1$-subdivision} of a graph $G$ is the graph obtained from $G$ by subdividing each edge of $G$ exactly once. We denote the $1$-subdivision of a star~$K_{1,s}$ by $K_{1,s}^1$;
in particular $K_{1,2}^1=P_5$.
Very recently, Chudnovsky, Spirkl and Zhong proved the following result:

\begin{theorem}[\cite{CSZ}]\label{t-known3}
For every $s\geq 1$ and $t\geq 1$, {\sc List $3$-Colouring} is polynomial-time solvable for $(K_{1,s}^1,P_t)$-free graphs. 
\end{theorem}

\noindent
For every $s\geq 1$ and $t\geq 2s+5$, the class of $(K_{1,{s+2}}^1,P_t)$-free graphs contains the class of $(sP_1+P_5)$-free graphs. Hence, Theorem~\ref{t-known3} generalizes Theorem~\ref{t-known} in the case $k=3$. 
As $K_{1,s}$ is an induced subgraph of $K_{1,s}^1$,
Theorem~\ref{t-known3} also generalizes the following result in the case $r=1$: 

\begin{theorem}[\cite{GPS14b}]\label{t-known4}
For every $k\geq 1$, $r\geq 1$, $s\geq 1$ and $t\geq 1$, {\sc List $k$-Colouring} is polynomial-time solvable for $(K_{r,s},P_t)$-free graphs. 
\end{theorem}

\subsection*{Our Results}

We prove the following result:

\begin{theorem}\label{t-new}
For every $r\geq 1$, $s\geq 1$ and $t\geq 1$, the mim-width of the class of $(K_r,K_{1,s}^1,P_t)$-free graphs is bounded and quickly computable.
\end{theorem}

\noindent
We may assume without loss of generality that an instance of {\sc List $k$-Colouring} is $K_{k+1}$-free, for otherwise it is a no-instance. Hence, combining Theorem~\ref{t-new} with Theorem~\ref{t-lc} enables us to generalize 
both Theorems~\ref{t-known} and~\ref{t-known3}:
 
 \begin{corollary}\label{c-new}
For every $k\geq 1$, $s\geq 1$ and $t\geq 1$, {\sc List $k$-Colouring} is polynomial-time solvable for $(K_{1,s}^1,P_t)$-free graphs. 
\end{corollary}

\noindent
Corollary~\ref{c-new} is tight in the following sense. Let $L_{1,s}$ denote the subgraph obtained from $K_{1,s}^1$ by subdividing one edge exactly once; in particular $L_{1,2}=P_6$.
Then, as {\sc List $4$-Colouring} is \NP-complete for $P_6$-free graphs~\cite{GPS14}, we cannot generalize Corollary~\ref{c-new} to $(L_{1,s},P_t)$-free graphs for $k \ge 4$, $s\geq 2$ and $t\geq 6$.
Moreover, the mim-width of $(K_4,P_6)$-free graphs is unbounded~\cite{BHMPP} and so we cannot extend Theorem~\ref{t-new} to $(K_r,L_{1,s},P_t)$-free graphs, 
for $r\geq 4$, $s\geq 2$ and $t\geq 6$, either. 

Theorem~\ref{t-new} has other applications as well. 
Firstly, as mentioned earlier, there are many problems known to be \XP\ parameterized by mim-width, so \cref{t-new} implies that these problems are polynomial-time solvable for this graph class; in particular, this is the case for the broad class of problems known as Locally Checkable Vertex Subset and Vertex Partitioning problems.
For a graph $G$, let $\omega(G)$ denote the size of a maximum clique in $G$.
Chudnovsky et al.~\cite{CKPR20} gave for the class of $(K_{1,3}^1,P_6)$-free graphs an $n^{O(\omega(G)^3)}$-time algorithm for \textsc{Max Partial $H$-Colouring}, a problem equivalent to {\sc Independent Set} if $H=P_1$ and to {\sc Odd Cycle Transversal} if $H=P_2$. 
In other words, \textsc{Max Partial $H$-Colouring} is polynomial-time solvable for  $(K_{1,3}^1,P_6)$-free graphs with bounded clique number. Moreover, they observed that \textsc{Max Partial $H$-Colouring} is polynomial-time solvable for graph classes whose mim-width is bounded and quickly computable. Hence, Theorem~\ref{t-new} generalizes their result for \textsc{Max Partial $H$-Colouring} to $(K_{1,s}^1,P_t)$-free graphs with bounded clique number, for any $s\geq 1$ and $t\geq 1$.
However, the running time of the corresponding algorithm is worse than $n^{O(\omega(G)^3)}$ (see~\cite{CKPR20} for details).

\smallskip
\noindent
It remains to prove Theorem~\ref{t-new}, which we do in the next section. In Section~\ref{s-con} we give some directions for future work. 

\section{The Proof of Theorem~\ref{t-new}}\label{s-new}

We first state two lemmas. The first lemma shows that given a partition of the vertex set of a graph $G$, we can bound the mim-width of $G$ in terms of the mim-width of the graphs induced by each part and the mim-width between any two of the parts.

\begin{lemma}\label{mimmultijoin}
Let $G$ be a graph, and let $(X_1,\dotsc,X_p)$ be a partition of $V(G)$ such that $\cutmim_G(X_i,X_j) \le c$ for all distinct $i,j \in \{1,\dotsc,p\}$, and $p \ge 2$.  Then \[\mimw(G) \le \max\left\{c\left\lfloor\left(\frac{p}{2}\right)^2\right\rfloor,\max_{i \in \{1,\dotsc,p\}}\{\mimw(G[X_i])\} + c(p-1)\right\}.\]
Moreover, if $(T_i,\delta_i)$ is a branch decomposition for $G[X_i]$ for each $i$, then we can construct, in $O(p)$ time, a branch decomposition $(T,\delta)$ for $G$ with $$\mimw(T,\delta) \le \max\left\{c\left\lfloor\left(\frac{p}{2}\right)^2\right\rfloor,\max_{i \in \{1,\dotsc,p\}}\{\mimw(T_i,\delta_i)\} + c(p-1)\right\}.$$
\end{lemma}

\begin{proof}
We construct a branch decomposition $(T,\delta)$ for $G$ with the desired mim-width as follows.
Let $T_0$ be an arbitrary subcubic tree having $p$ leaves $\ell_1,\dotsc,\ell_p$.
Fix for each $i \in \{1,\ldots,p\}$ a branch decomposition $(T_i,\delta_i)$ for $G[X_i]$.
For each $i \in \{1,\dotsc,p\}$, we choose an arbitrary leaf vertex $v_i$ of $T_i$, we identify $v_i$ with $\ell_i$ calling the resulting vertex $\ell_i$, and we create a new pendant edge incident to $\ell_i$, where the new leaf vertex adjacent to $\ell_i$ is called $v_i$.
Then $T$ is a subcubic tree whose set of leaves is the disjoint union of the leaves of $T_i$ for each $i \in \{1,\dotsc,p\}$. 
See \cref{trees-fig}, for example.
For a leaf $v$ of $T$, we set $\delta(v) = \delta_i(v)$, where $v$ is a leaf of $T_i$.
Now $(T,\delta)$ is a branch decomposition for $G$, and clearly this branch decomposition can be constructed in $O(p)$ time.
It remains to prove the upper bound for $\mimw(T,\delta)$.

\begin{figure}
  \centering
  \begin{tikzpicture}[scale=0.49,line width=1pt]
    \tikzset{VertexStyle/.append style = {minimum height=5,minimum width=5}}
    \node at (1,-1.5) {\large$T_0$};

    \Vertex[x=-1,y=1,LabelOut=true,L=$\ell_1$,Lpos=180]{l1}
    \Vertex[x=-1,y=-1,LabelOut=true,L=$\ell_2$,Lpos=180]{l2}
    \Vertex[x=3,y=1,LabelOut=true,L=$\ell_{p-1}$,Lpos=0]{l3}
    \Vertex[x=3,y=-1,LabelOut=true,L=$\ell_{p}$,Lpos=0]{l4}

    \SetVertexNoLabel
    \tikzset{VertexStyle/.append style = {fill=gray}}
    \Vertex[x=0,y=0]{i0}
    \Vertex[x=2,y=0]{i1}

    \Edge(i0)(l1)
    \Edge(i0)(l2)
    \tikzset{EdgeStyle/.append style = {dotted}}
    \Edge(i0)(i1)
    \tikzset{EdgeStyle/.append style = {solid}}
    \Edge(i1)(l3)
    \Edge(i1)(l4)

    \node at (7.0,2) {\large$T_1$};

    \tikzset{VertexStyle/.append style = {fill=black}}
    \Vertex[x=8,y=3]{t1_1}
    \Vertex[x=8,y=1]{t1_2}
    \Vertex[x=12,y=3]{t1_3}
    \SetVertexLabel
    \Vertex[x=12,y=1,LabelOut=true,L=$v_1$,Lpos=0]{t1_4}
    \SetVertexNoLabel

    \tikzset{VertexStyle/.append style = {fill=gray}}
    \Vertex[x=9,y=2]{t1_i0}
    \Vertex[x=11,y=2]{t1_i1}

    \Edge(t1_i0)(t1_1)
    \Edge(t1_i0)(t1_2)
    \Edge(t1_i0)(t1_i1)
    \Edge(t1_i1)(t1_3)
    \Edge(t1_i1)(t1_4)

    \node at (7.0,-2) {\large$T_2$};

    \tikzset{VertexStyle/.append style = {fill=black}}
    \Vertex[x=8,y=-3]{t2_1}
    \Vertex[x=8,y=-1]{t2_2}
    \Vertex[x=12,y=-3]{t2_3}
    \Vertex[x=10,y=-3.3]{t2_5}
    \SetVertexLabel
    \Vertex[x=12,y=-1,LabelOut=true,L=$v_2$,Lpos=0]{t2_4}
    \SetVertexNoLabel

    \tikzset{VertexStyle/.append style = {fill=gray}}
    \Vertex[x=9,y=-2]{t2_i0}
    \Vertex[x=10,y=-2]{t2_i2}
    \Vertex[x=11,y=-2]{t2_i1}

    \Edge(t2_i0)(t2_1)
    \Edge(t2_i0)(t2_2)
    \Edge(t2_i0)(t2_i2)
    \Edge(t2_i2)(t2_i1)
    \Edge(t2_i2)(t2_5)
    \Edge(t2_i1)(t2_3)
    \Edge(t2_i1)(t2_4)

    \node at (20.0,2.5) {\large$T_{p-1}$};

    \tikzset{VertexStyle/.append style = {fill=black}}
    \Vertex[x=18,y=3.3]{t3_1}
    \SetVertexLabel
    \Vertex[x=17,y=1,LabelOut=true,L=$v_{p-1}$,Lpos=180]{t3_2}
    \SetVertexNoLabel
    \Vertex[x=19,y=1]{t3_4}

    \tikzset{VertexStyle/.append style = {fill=gray}}
    \Vertex[x=18,y=2]{t3_i0}

    \Edge(t3_i0)(t3_1)
    \Edge(t3_i0)(t3_2)
    \Edge(t3_i0)(t3_4)

    \node at (21.0,-2) {\large$T_p$};

    \tikzset{VertexStyle/.append style = {fill=black}}
    \Vertex[x=16,y=-3]{t4_1}
    \SetVertexLabel
    \Vertex[x=16,y=-1,LabelOut=true,L=$v_p$,Lpos=180]{t4_2}
    \SetVertexNoLabel
    \Vertex[x=20,y=-3]{t4_3}
    \Vertex[x=20,y=-1]{t4_4}
    \Vertex[x=17,y=-4.3]{t4_5}
    \Vertex[x=19,y=-4.3]{t4_6}

    \tikzset{VertexStyle/.append style = {fill=gray}}
    \Vertex[x=17,y=-2]{t4_i0}
    \Vertex[x=19,y=-2]{t4_i1}
    \Vertex[x=18,y=-2]{t4_i2}
    \Vertex[x=18,y=-3.3]{t4_i3}

    \Edge(t4_i0)(t4_1)
    \Edge(t4_i0)(t4_2)
    \Edge(t4_i0)(t4_i2)
    \Edge(t4_i2)(t4_i1)
    \Edge(t4_i1)(t4_3)
    \Edge(t4_i1)(t4_4)
    \Edge(t4_i2)(t4_i3)
    \Edge(t4_i3)(t4_5)
    \Edge(t4_i3)(t4_6)
  \end{tikzpicture}\\[15pt]
    \begin{tikzpicture}[scale=0.49]
        \tikzset{VertexStyle/.append style = {minimum height=5,minimum width=5}}
    \node at (28.0,-5) {\large$T$};

    \SetVertexLabel
    \tikzset{VertexStyle/.append style = {fill=white}}
    \Vertex[x=28,y=1,LabelOut=true,L=$\ell_1$,Lpos=180]{t1}
    \Vertex[x=28,y=-1,LabelOut=true,L=$\ell_2$,Lpos=180]{t2}
    \Vertex[x=34,y=1.5,LabelOut=true,L=$\ell_{p-1}$,Lpos=-90]{t3}
    \Vertex[x=33,y=-1,LabelOut=true,L=$\ell_{p}$,Lpos=0]{t4}

    \SetVertexNoLabel
    \tikzset{VertexStyle/.append style = {fill=gray}}
    \Vertex[x=29,y=0]{i0}
    \Vertex[x=32,y=0]{i1}

    \Edge(i0)(t1)
    \Edge(i0)(t2)
    \tikzset{EdgeStyle/.append style = {dotted}}
    \Edge(i0)(i1)
    \tikzset{EdgeStyle/.append style = {solid}}
    \Edge(i1)(t3)
    \Edge(i1)(t4)

    \tikzset{VertexStyle/.append style = {fill=black}}
    \Vertex[x=24,y=3]{t_1_1}
    \Vertex[x=24,y=1]{t_1_2}
    \Vertex[x=28,y=3]{t_1_3}
    \SetVertexLabel
    \Vertex[x=29,y=2,LabelOut=true,L=$v_1$,Lpos=0]{t_1_4}
    \SetVertexNoLabel

    \tikzset{VertexStyle/.append style = {fill=gray}}
    \Vertex[x=25,y=2]{t_1_i0}
    \Vertex[x=27,y=2]{t_1_i1}

    \Edge(t_1_i0)(t_1_1)
    \Edge(t_1_i0)(t_1_2)
    \Edge(t_1_i0)(t_1_i1)
    \Edge(t_1_i1)(t_1_3)
    \Edge(t_1_i1)(t1)
    \Edge(t1)(t_1_4)

    \tikzset{VertexStyle/.append style = {fill=black}}
    \Vertex[x=24,y=-3]{t_2_1}
    \Vertex[x=24,y=-1]{t_2_2}
    \Vertex[x=28,y=-3]{t_2_3}
    \Vertex[x=26,y=-3.3]{t_2_5}
    \SetVertexLabel
    \Vertex[x=29,y=-2,LabelOut=true,L=$v_2$,Lpos=-45]{t_2_4}
    \SetVertexNoLabel

    \tikzset{VertexStyle/.append style = {fill=gray}}
    \Vertex[x=25,y=-2]{t_2_i0}
    \Vertex[x=26,y=-2]{t_2_i2}
    \Vertex[x=27,y=-2]{t_2_i1}

    \Edge(t_2_i0)(t_2_1)
    \Edge(t_2_i0)(t_2_2)
    \Edge(t_2_i0)(t_2_i2)
    \Edge(t_2_i2)(t_2_i1)
    \Edge(t_2_i2)(t_2_5)
    \Edge(t_2_i1)(t_2_3)
    \Edge(t_2_i1)(t2)
    \Edge(t2)(t_2_4)

    \tikzset{VertexStyle/.append style = {fill=black}}
    \Vertex[x=35,y=3.8]{t_3_1}
    \SetVertexLabel
    \Vertex[x=33,y=2.5,LabelOut=true,L=$v_{p-1}$,Lpos=135]{t_3_2}
    \SetVertexNoLabel
    \Vertex[x=36,y=1.5]{t_3_4}

    \tikzset{VertexStyle/.append style = {fill=gray}}
    \Vertex[x=35,y=2.5]{t_3_i0}

    \Edge(t_3_i0)(t_3_1)
    \Edge(t_3_i0)(t3)
    \Edge(t3)(t_3_2)
    \Edge(t_3_i0)(t_3_4)

    \tikzset{VertexStyle/.append style = {fill=black}}
    \Vertex[x=33,y=-3]{t_4_1}
    \SetVertexLabel
    \Vertex[x=32,y=-2,LabelOut=true,L=$v_p$,Lpos=180]{t_4_2}
    \SetVertexNoLabel
    \Vertex[x=37,y=-3]{t_4_3}
    \Vertex[x=37,y=-1]{t_4_4}
    \Vertex[x=34,y=-4.3]{t_4_5}
    \Vertex[x=36,y=-4.3]{t_4_6}

    \tikzset{VertexStyle/.append style = {fill=gray}}
    \Vertex[x=34,y=-2]{t_4_i0}
    \Vertex[x=36,y=-2]{t_4_i1}
    \Vertex[x=35,y=-2]{t_4_i2}
    \Vertex[x=35,y=-3.3]{t_4_i3}

    \Edge(t_4_i0)(t_4_1)
    \Edge(t_4_i0)(t4)
    \Edge(t4)(t_4_2)
    \Edge(t_4_i0)(t_4_i2)
    \Edge(t_4_i2)(t_4_i1)
    \Edge(t_4_i1)(t_4_3)
    \Edge(t_4_i1)(t_4_4)
    \Edge(t_4_i2)(t_4_i3)
    \Edge(t_4_i3)(t_4_5)
    \Edge(t_4_i3)(t_4_6)
  \end{tikzpicture}
  \caption{An example of the construction of $T$ in the proof of \cref{mimmultijoin}.}
  \label{trees-fig}
\end{figure}
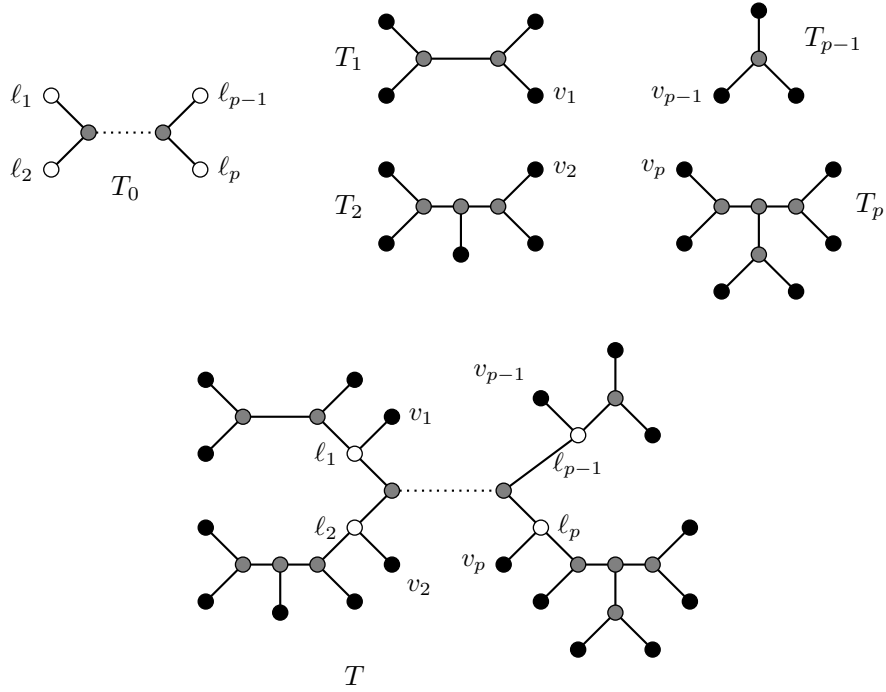

Consider $e \in E(T)$ and the partition $(A_e,\overline{A_e})$ of $V(G)$.
If $e \in E(T_0)$, then $A_e = \bigcup_{j \in J} X_j$ for some $J \subseteq \{1,\dots,p\}$.
If $e \in E(T_i)$ for some $i \in \{1,\dotsc,p\}$, then either $A_e$ or $\overline{A_e}$ is properly contained in $X_i$.
The only other possibility is that $e$ is one of the newly created pendant edges, in which case either $A_e$ or $\overline{A_e}$ has size~$1$.

First suppose $e\in E(T_0)$, so $A_e = \bigcup_{j \in J} X_j$ for some $J \subseteq \{1,\dots,p\}$.
We claim that $\cutmim_G(A_e,\overline{A_e}) \le c\left\lfloor\left(\frac{p}{2}\right)^2\right\rfloor$.
Let $M$ be a maximum-sized induced matching in $G[A_e,\overline{A_e}]$.
Let $K = \{1,\dotsc,p\} \setminus J$.
For each $j \in J$ and $k \in K$, there are at most $c$ edges of $M$ with one end in $X_j$ and the other end in $X_k$, since $\cutmim_G(X_j,X_k) \le c$.
Thus $\cutmim_G(A_e,\overline{A_e}) \le c|J||K|$, where $|J| + |K| = p$.
As $c|J||K| \le c\left\lfloor\left(\frac{p}{2}\right)\right\rfloor\left\lceil\left(\frac{p}{2}\right)\right\rceil = c\left\lfloor\left(\frac{p}{2}\right)^2\right\rfloor$, the claim follows.

Now suppose $e \in E(T_i)$ for some $i \in \{1,\dotsc,p\}$, so, without loss of generality, $A_e$ is properly contained in $X_i$.
We claim that $\cutmim_G(A_e,\overline{A_e}) \le \mimw(G[X_i]) + c(p-1)$.
Consider a maximum-sized induced matching $M$ in $G[A_e,\overline{A_e}]$.
As $A_e \subseteq X_i$, all the edges of $M$ have one end in $X_i$.
For each $j \in \{1,\dotsc,p\}$ with $j \neq i$, there are at most $c$ edges of $M$ with one end in $X_j$, since $\cutmim_G(X_i,X_j) \le c$.
Since there are at most $\mimw(G[X_i])$ edges of $M$ with both ends in $X_i$, we deduce that $\cutmim_G(A_e,\overline{A_e}) \le \mimw(G[X_i]) + c(p-1)$, as claimed.  The lemma follows.
\end{proof}

A {\it clique} in a graph is a set of pairwise adjacent vertices. An {\it independent set} is a set of pairwise non-adjacent vertices.
A {\it dominating set} is a set $D$ of vertices such that every vertex not in $D$ is adjacent to at least one vertex in $D$.
Ramsey's Theorem states that for all positive integers $k$ and $\ell$, there exists an integer $R(k,\ell)$ such that every graph on at least $R(k,\ell)$ vertices contains a clique of size~$k$ or an independent set of size~$\ell$.
A well-known, rough bound for $R(k,\ell)$
is $R(k,\ell) \le \binom{k + \ell - 2}{k-1} \le (k+\ell-2)^{k-1}$.

For $r\geq 1$ and $s,t\geq 1$, let $M(r,s,t)=(1+R(r+1,R(r+1,s)))^{t-2}$.
The next lemma has been proven by Chudnovsky, Spirkl and Zhong~\cite{CSZ} for the case where $r=3$. The proof of the lemma is analogous to the proof in~\cite{CSZ} for the case where $r=3$: replace each occurrence of ``$4$'' in the proofs of 
Lemmas~13 and~15 in~\cite{CSZ} by ``$r+1$''.

\begin{lemma}[cf.~\cite{CSZ}]\label{l-dom}
For every $r\geq 1$, $s\geq 1$ and $t\geq 1$, a connected 
$(K_{r+1},K_{1,s}^1,P_t)$-free 
graph contains a dominating set of size at most $M(r,s,t)$.
\end{lemma}

\noindent
We are now ready to prove Theorem~\ref{t-new}.  We in fact prove the following theorem, \cref{t-new-detailed}, which gives an explicit bound on the mim-width; \cref{t-new} then follows from this. 

\begin{theorem}
  \label{t-new-detailed}
Let $r\geq 1$, $s\geq 1$ and $t\geq 1$, and let $G$ be a $(K_r,K_{1,s}^1,P_t)$-free graph.
Then $\mimw(G) \le g(r,s,t)$ where $g(r,s,t)=2(r+s-1)^{2(r+1)^2(t+1)}$, and a branch decomposition $(T,\delta)$ of $G$ with $\mimw(T,\delta) \le g(r,s,t)$ can be found in polynomial time.
\end{theorem}

\begin{proof}
We may assume without loss of generality that $G$ is connected.
We use induction on~$r$. If $r\leq 2$, then $G$ is $K_2$-free, so $\mimw(T,\delta)=0$ for any branch decomposition $(T,\delta)$ of $G$, whereas $g(r,s,t)$ is positive for all $s,t \ge 1$; so the theorem holds trivially in this case.

Suppose that $r\geq 3$. 
By Lemma~\ref{l-dom}, we find that $G$ has a dominating set $D$ of size at most 
$M(r-1,s,t)$. 
Moreover, we can find $D$ in polynomial time by brute force (or we can apply the $O(tn^2)$-time algorithm of~\cite{CSZ}).
We let $p=|D|$, so $p \le M(r-1,s,t)$.

Let $f(r,s,t) = (r+s-1)^{2(r+1)^2(t+1)}$.
We will show that there is a branch decomposition $(T',\delta')$ of $G-D$ with $\mimw(T',\delta') \le f(r,s,t)$.
The theorem will then follow: to see this, observe that if $(T',\delta')$ is such a branch decomposition, then we can readily extend $(T',\delta')$ to a branch decomposition $(T,\delta)$ for $G$ with mim-width at most $f(r,s,t) + p \le f(r,s,t) + M(r-1,s,t) \le g(r,s,t)$.
Namely, we can obtain $T$ in polynomial time from $T'$ and an arbitrary subcubic tree $T''$ with $p+2$ leaves by identifying a leaf of $T'$ with a leaf of $T''$.
So it remains to prove that $\mimw(G-D) \le f(r,s,t)$, and that we can find a branch decomposition witnessing this bound, in polynomial time.

Let $V=V(G)$.
We partition $V$ 
as follows.
We first fix an arbitrary ordering $d_1,\ldots,d_p$ on the vertices of $D$.
Let $X_1$ be the set of vertices in $V\setminus D$ adjacent to $d_1$.
For $i\in \{2,\ldots,p\}$, let $X_i$ be the set of vertices in $V\setminus D$
adjacent to $d_i$, but non-adjacent to any $d_h$ with $h\leq i-1$. 
Then $\{D,X_1,\ldots,X_p\}$ is a partition of $V$ (where some of the sets $X_i$ might be empty). 
Moreover, we found this partition in polynomial time.

By construction, $d_i$ is adjacent to every vertex of $X_i$ for each $i\in \{1,\ldots,p\}$. As $G$ is $K_r$-free, this implies that each $X_i$ induces a $(K_{r-1},K_{1,s}^1,P_t)$-free subgraph of $G$.
By the induction hypothesis, $\mimw(G[X_i]) \le f(r-1,s,t) + M(r-2,s,t)$, 
and a branch decomposition witnessing this mim-width bound can be computed in polynomial time, 
for every $i\in \{1,\ldots,p\}$.

Consider two sets $X_i$ and $X_j$ with $i<j$. 
We claim that $\cutmim_G(X_i,X_j) < c=R(r-1,R(r-1,s))$.
Towards a contradiction, suppose that $\cutmim_G(X_i,X_j) \geq c$. 
Then, by definition, there exist two sets $A= \{a_1,a_2,\dotsc,a_c\} \subseteq X_i$ and $B =
\{b_1,b_2,\dotsc,b_c\} \subseteq X_j$, each of size $c$, such that $\{a_1b_1,\ldots,a_cb_c\}$ is a set of $c$ edges with the property that $G$ does not contain any edges $a_ib_j$ for $i\neq j$ (note that edges $a_ia_j$ and $b_ib_j$ may exist in $G$).

As $G[X_i]$ is $K_{r-1}$-free, and $|A| = c=R(r-1,R(r-1,s))$, Ramsey's Theorem tells us that $G[A]$ contains an independent set $A'$ of size $c'=R(r-1,s)$. Assume without loss of generality that $A'=\{a_1,\ldots,a_{c'}\}$. Let $B'=\{b_1,\ldots,b_{c'}\}$. As $G[X_j]$ is $K_{r-1}$-free, $G[B']$ contains an independent set $B''$ of size~$s$. Assume without loss of generality that $B''=\{b_1,\ldots,b_s\}$. 
By construction, $d_i$ is adjacent to every vertex of $\{a_1,\ldots,a_s\}\subseteq X_i$ and non-adjacent to every vertex of $\{b_1,\ldots,b_s\}\subseteq X_j$.
Hence, $\{a_1,\ldots,a_s,b_1,\ldots,b_s,d_i\}$ induces a $K_{1,s}^1$ in $G$, a contradiction. 
We conclude that $\cutmim_G(X_i,X_j) < c$.

Now, by \cref{mimmultijoin}, we have 
\[\begin{array}{lcl}
\mimw(G-D) &\le &\max\left\{c\left\lfloor\left(\frac{p}{2}\right)^2\right\rfloor,\max_{i \in \{1,\dotsc,p\}}\{\mimw(G[X_i])\} + c(p-1)\right\}\\[8pt]
& \le &\max\left\{ cp^2,\,f(r-1,s,t) + M(r-2,s,t)+cp\right\}.\end{array}\] 
Recall that  $R(k,\ell) \le (k+\ell-2)^{k-1}$. We observe that $R(k,R(k,\ell)) \le (k+\ell-2)^{k(k-1)}$. 
Hence,  $c=R(r-1,R(r-1,s))\le  (r+s-3)^{(r-1)(r-2)}$ and $p \le M(r-1,s,t) = (1+R(r,R(r,s)))^{t-2} \le \left(1+(r+s-1)^{r(r+1)}\right)^{t-2} \le \left((r+s-1)^{r(r+1)+1}\right)^{t-2}$.
Thus
\begin{align*}
  cp^2
  &\le (r+s-3)^{(r-1)(r-2)} \left((r+s-1)^{r(r+1)+1}\right)^{2(t-2)} \\
  &\le (r+s-1)^{(r+1)^2} (r+s-1)^{2(r+1)^{2}(t-2)} 
  \le (r+s-1)^{2(r+1)^{2}t} \le f(r,s,t), \textrm{ and}
\end{align*}

\[\begin{array}{lcl} 
&&f(r-1,s,t) + M(r-2,s,t)+cp\\[8pt]  
  &\le & (r+s-2)^{2r^2(t+1)} + \left((r+s-2)^{(r-1)r+1}\right)^{t-2} + (r+s-3)^{(r-1)(r-2)} \left((r+s-1)^{r(r+1)+1}\right)^{t-2} \\[8pt]
  &\le &(r+s-1)^{r^2(t+1)}\left((r+s-1)^{r^2(t+1)} + 1 + (r+s-1)^{(r+1)(t-1)} \right) \\[8pt]
  &\le &(r+s-1)^{r^2(t+1)}\left((r+s-1)^{r^2(t+1)+1}\right)\\[8pt] 
  &= &(r+s-1)^{2r^2(t+1)+1}\\[8pt] &\le &f(r,s,t).
\end{array}\]

\noindent
So $\mimw(G-D) \le f(r,s,t)$ and the theorem follows by induction.
\end{proof}

\section{Conclusions}\label{s-con}

We proved in Corollary~\ref{c-new} that for every $k\geq 1$, $s\geq 1$ and $t\geq 1$, {\sc List $k$-Colouring} is polynomial-time solvable for $(K_{1,s}^1,P_t)$-free graphs by showing that the mim-width of these graphs is bounded and quickly computable. Huang~\cite{Hu16} proved that {\sc $4$-Colouring} is \NP-complete for $P_7$-free graphs and that {\sc $5$-Colouring} is \NP-complete for $P_6$-free graphs. It is also known that {\sc List $4$-Colouring} is \NP-complete for $P_6$-free graphs~\cite{GPS14}.
However, the {\sc List $3$-Colouring} problem is polynomial-time solvable for $P_7$-free graphs~\cite{BCMSSZ18} and the computational complexities of {\sc $3$-Colouring} and {\sc List $3$-Colouring} are open for $P_t$-free graphs if $t\geq 8$. In particular, we do not know any integer $t$ such that {\sc $3$-Colouring} or {\sc List $3$-Colouring} are \NP-complete for $P_t$-free graphs. Recently, Pilipczuk, Pilipczuk and Rz{\k{a}}\.{z}ewski~\cite{PPR} gave, for every $t\geq 3$,
a quasi-polynomial-time algorithm for {\sc $3$-Colouring} on the class of $\{C_{t+1},C_{t+2},\ldots\}$-free graphs; note that this class contains, for $t\geq 2$, the class of $P_t$-free graphs as a subclass.
Hence, an extension of Corollary~\ref{c-new}, which will require more research into the structure of $P_t$-free graphs, might still be possible for $k=3$. We leave this for future work.

\end{document}